%% file: main.tex
\documentclass[conference]{IEEEtran}
\IEEEoverridecommandlockouts
\usepackage{cite}
\usepackage{amsmath,amssymb,amsfonts}
\usepackage{graphicx}
\usepackage{textcomp}
\usepackage{xcolor}
\usepackage{bbm}
\usepackage[english]{babel}
\usepackage{algorithm}
\usepackage[utf8]{inputenc}
\usepackage[noend]{algpseudocode}
\usepackage{amsmath,amssymb,amsfonts}
\usepackage{verbatim}
\usepackage{amsmath}
\usepackage{amsmath,amssymb,amsthm}
\usepackage{array,booktabs}
\usepackage{multirow}
\usepackage{graphicx}
\usepackage{textcomp}
\usepackage{soul,xcolor}
\usepackage{comment}
\usepackage{multirow}
\usepackage{color, colortbl}
\usepackage{subfig}
\usepackage{hyperref}
\usepackage{float}
\usepackage{nomencl}
\usepackage{wasysym}
\usepackage{soul}
\usepackage{color}
\usepackage[colorinlistoftodos]{todonotes}
\usepackage{nomencl}
\addto\captionsenglish{\renewcommand{\figurename}{Fig.}}

\usepackage{tikz}

\newtheorem{lemma}{\bf{Lemma}}
\newtheorem{definition}{{Definition}}

\begin{document}

\title{Chaos Engineering for Enhanced Resilience of Cyber-Physical Systems}

\author{Charalambos~Konstantinou,~\IEEEmembership{Senior~Member,~IEEE}, 
George Stergiopoulos,~\IEEEmembership{Member,~IEEE},\protect \\ Masood Parvania,~\IEEEmembership{Senior~Member,~IEEE}, and Paulo Esteves-Verissimo,~\IEEEmembership{Fellow,~IEEE}

\thanks{C. Konstantinou and P. Esteves-Verissimo are with the Computer, Electrical and Mathematical Sciences and Engineering Division, King Abdullah University of Science and Technology (KAUST),  Thuwal 23955-6900, Saudi Arabia (e-mail: {\{\href{mailto:charalambos.konstantinou@kaust.edu.sa}{charalambos.konstantinou}, \href{mailto:paulo.verissimo@kaust.edu.sa}{paulo.verissimo}\}@kaust.edu.sa)}. \protect

G. Stergiopoulos is with the Department of Information and Communication Systems Engineering, University of the Aegean, Greece (e-mail: \href{mailto:g.stergiopoulos@aegean.gr}{g.stergiopoulos@aegean.gr}).
\protect 

M. Parvania is  with the Department of Electrical and Computer Engineering, University of Utah, Salt Lake
City, UT 84112 USA (e-mail: \href{mailto:masood.parvania@utah.edu}{masood.parvania@utah.edu}).
}
}

\maketitle

\begin{abstract}

Cyber-physical systems (CPS) incorporate the complex and large-scale engineered systems behind critical infrastructure operations, such as water distribution networks, energy delivery systems, healthcare services, manufacturing systems, and transportation networks. 
Industrial CPS in particular need to simultaneously satisfy requirements of available, secure, safe and reliable system operation against diverse threats, in an adaptive and sustainable way. 
These adverse events can be of accidental or malicious nature and may include natural disasters, hardware or software faults, cyberattacks, or even infrastructure design and implementation faults.  They may drastically affect the results of CPS algorithms and mechanisms, and subsequently the operations of industrial control systems (ICS) deployed in those critical infrastructures. 
Such a demanding combination of properties and threats calls for resilience-enhancement  methodologies and techniques, working in real-time operation. However, the analysis of CPS resilience is a difficult task as it involves evaluation of various interdependent layers with heterogeneous computing equipment, physical components, network technologies, and data analytics. 
In this paper, we apply the principles of chaos engineering (CE) to industrial CPS, in order to demonstrate the benefits of such practices on system resilience. 
The systemic uncertainty of adverse events can be tamed by applying runtime CE-based analyses to CPS in production, in order to predict environment changes and thus apply mitigation measures limiting the range and severity of the event, and minimizing its blast radius.

\end{abstract}


\input{1-introduction_PJV}

\input{2-literature_PJV}
\input{3-cemodel_PJV}
\input{4-usecase}

\input{5-conclusions_PJV}

\bibliographystyle{IEEEtran}
\bibliography{biblio}

\end{document}

%% file: 1-introduction_PJV.tex
\section{Introduction} \label{Introduction}

The term cyber-physical systems (CPS) refers to systems which integrate and interconnect the physical environment with control, computing, and networking components. In the last decade, there has been a rapid growth of CPS in critical infrastructures including power systems, manufacturing facilities, robotics, transportation networks, desalination plants, and other industrial control systems (ICS). 

These substantial applications of CPS, especially in industrial environments, have largely computerized critical infrastructures, making them subject to diverse threats which industrial CPS need to face, by simultaneously satisfying requirements of available, secure, safe and reliable system operation, in an adaptive and sustainable way. 

These adverse events can be of accidental or malicious nature and may include natural disasters such as hurricanes, earthquakes, wildfires, hardware or software faults, e.g., in smart monitoring devices due to bugs in the code (e.g., operating system, compilers, libraries, etc.), or bit flips induced by cosmic rays, or ‘silent errors’ due to components and manufacturing variability faults, as well as cyberattacks, infrastructure design and implementation faults and, last but not least, unexpected operating conditions. They may drastically affect the results of CPS algorithms and mechanisms, and subsequently the operations of ICS deployed in those critical infrastructures, leading in the end to potential financial, social, legal, and political consequences. 

Such a demanding combination of requirements versus threats calls for correspondingly powerful defenses, beyond classic industry-practice mitigation techniques.  
In other words, industrial CPS should seek \emph{resilience}, or the continued ability to  ``endure'' and ``recover'' from disrupting events. Resilient computing systems are ones that: have built-in baseline defences; can cope with virtually any quality of threat, be it accidental or malicious; 
protect in an incremental way and automatically respond and adapt to threats; and provide unattended and sustainable operation \cite{paulo1}. However, the cratfing and analysis of CPS resilience is a difficult task as it involves, as usual in this sector,  various interdependent layers with heterogeneous computing equipment, physical components, network technologies, and data analytics.

\begin{figure}
    \centering
    \includegraphics[width=1\columnwidth]{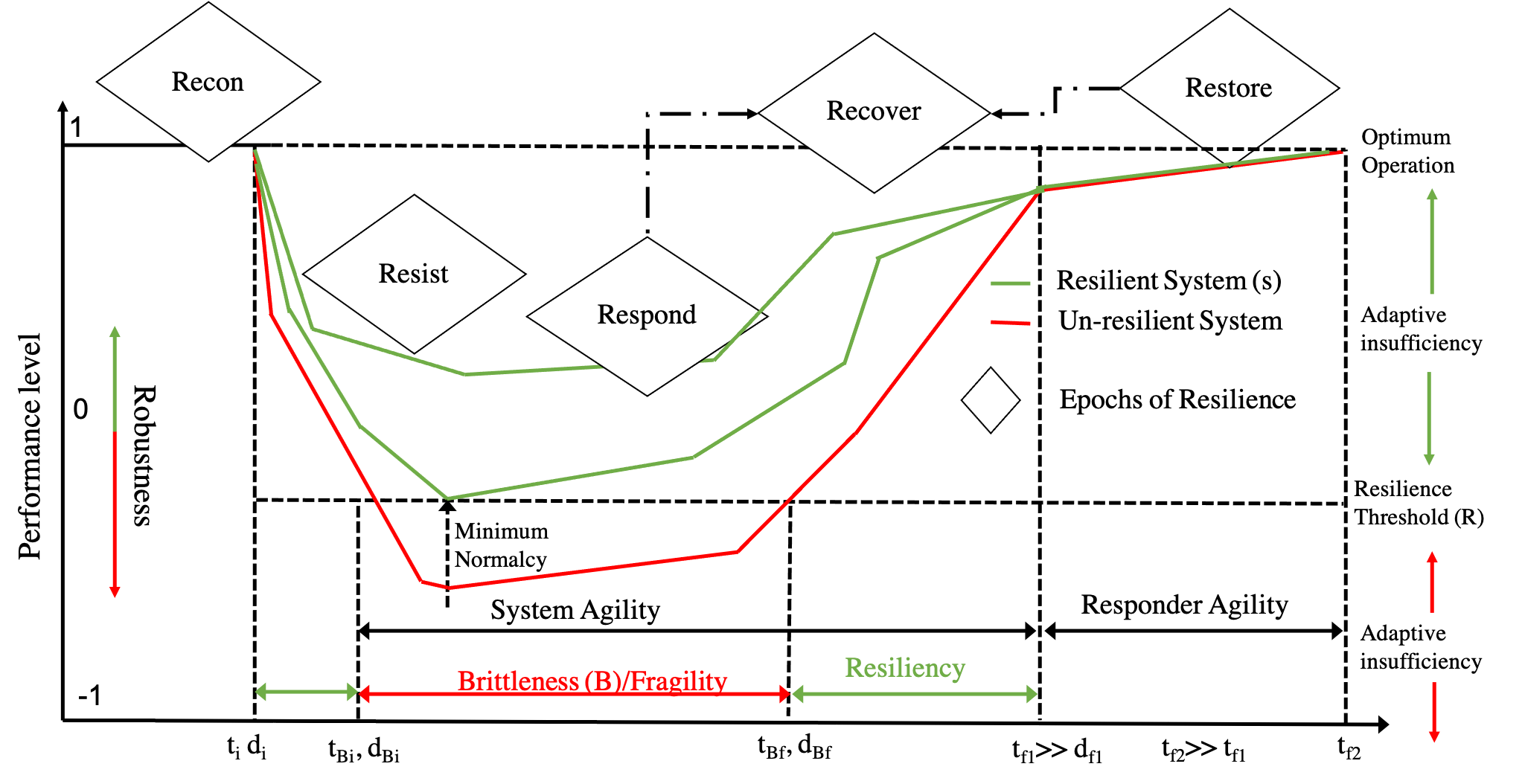}
    \caption{Disturbance and impact resilience evaluation (DIRE) Curve ($i$=initial, $f$=final) \cite{8088656}.}
    \label{fig:dire}
\end{figure}

In order to enhance and evaluate CPS resilience, there exist various works on resilience definitions, metrics, and evaluation and enhancement methods. A conceptual resilience curve with resilient control metrics is the ``Disturbance and Impact Resilience Evaluation Curve'' (DIRE), presented in Fig. \ref{fig:dire}, expressing the ``Rs'' of resilience, i.e., Recon, Resist, Respond, Recover, and Restore according to \cite{8088656}. Experimentation practices involve the use of hardware-in-the-loop (HIL) testbed environments \cite{9351954} and digital twins \cite{koulamas2018cyber}, Monte Carlo simulation-based methods \cite{yu2020resilient}, pre-selected scenario-based methods \cite{9496085}, and machine learning-based methods \cite{8255544}. 
Existing methods, however, are often not examined in deployed CPS nor governed by a proactive experimentation practice in which weaknesses can be revealed before impacting the system, e.g., often HIL experimentation is specific to the application and scenarios under test \cite{8221708}, and not following systematic experimentation steps. 


In this work, we apply the principles of chaos engineering (CE) to industrial CPS, in order to demonstrate the benefits of such practices on system resilience. CE has been primarily used for software systems, and can be defined as ``the discipline of experimenting on a system in order to build confidence in the system’s capability to withstand turbulent conditions in production'' \cite{miles2019learning}.
We propose to extend the concept to a more general framework including overall CPS architectures (hardware and software), as well as a broader scope of faults beyond software.

\begin{figure}
    \centering
    \includegraphics[width=\columnwidth]{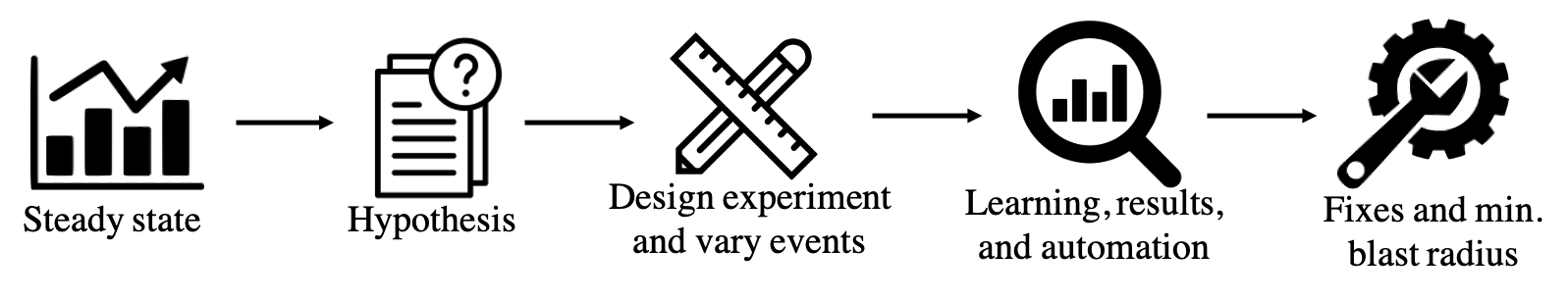}
    \caption{Chaos engineering (CE) experimentation principles.}
    \label{fig:steps}
\end{figure}

CE involves the facilitation of experiments to reveal weaknesses within a system, i.e., how the system responds to unexpected events such as outages due to cyberattacks or faults. The concept of CE was originated from Netflix in an effort to develop a built-in resilience tool within the production environment \cite{basiri2016chaos}. Typically, CE experiments include the following steps (Fig. \ref{fig:steps}): (1) define the `steady state' of a system in order to quantify its characteristic behavior under a certain type of events; (2) build a hypothesis around the steady state which should be followed in both the control/theoretical aspect and the experiments (measurable output of a system); (3) vary process events into the systems reflecting realistic adverse productions conditions; (4) run automated experiments within the system production environment in an effort to disprove the hypothesis in terms of control and experiments differences. 

This experimentation process builds confidence in the system with regards to its steady states of operation and how hard they are to be disrupted. In case CE experimentation identifies a systemic weakness, the overall process is set up in a way to ensure that potential impact radius from the experiment is minimized and contained. The systemic uncertainty of adverse events can thus be tamed by applying runtime CE-based analyses to CPS in production, in order to predict environment changes and thus apply mitigation measures limiting the range and severity of the event, and minimizing its blast radius\footnote{Blast radius is often defined what is impacted with the CE tests and experiments. The effort to ``minimize the blast radius'' essentially refers to effort of identifying via CE experimentation the vulnerabilities causing failure ``without accidentally blowing everything up'' \cite{rosenthal2017chaos}.}.




In terms of quantitative benefits, CE can reduce the number of preventable outages. This reduction of the blast radius is coupled with chaos-induced bounded harm that is revealed by CE experiments. 
Another aiding approach to CE is modeling outages using historical data, i.e., a proper implementation of CE experiments requires an establishment of a baseline for test metrics based on historical data. Taking proper records from an incident management system in place, and identifying vulnerabilities where the failure of a seemingly low-risk node (i.e., with subliminal criticality) could bring the system down leading to an outage, provides insight to quantitative financial loss due to the occurred incident as well as information about vulnerable nodes that should be injected in the next CE experiment. The adoption process may require engineers to mitigate risks of impacting customers with these experiments. This means there is a balance between reducing the impact to customers and identifying vulnerabilities quickly. As for qualitative benefits of CE, these include improvement in design decisions, growth in understanding service criticality, and confidence in validating reliability measures.


In this paper, we introduce the concept of CE for enhancing the resilience of CPS, and specifically, ICS environments of critical infrastructures. First, we review experimentation practices for weaknesses identification and testing as well as present existing CE approaches (Section \ref{s:related}). We then build a case of a CE formal model in which we define and describe the control theoretical aspects of CE within a nonlinear cyber-physical ICS (Section \ref{s:model}). Furthermore, we present a use-case scenario on the industrial process control model of the Tennessee Eastman Challenge in an effort to map the CE experimentation steps with actual metrics and failure events (Section \ref{s:use-case}). Finally, Section \ref{s:conclusions} concludes the paper and provides directions for  future work.

%% file: 2-literature_PJV.tex
\section{Related Work and Limitations}\label{s:related} 

Modern distributed systems and content delivery networks are often implemented as software-based services. Testing and vulnerability detection in distributed systems is a broad field that borrows numerous techniques from different engineering fields. CPS, and in particular modern ICS, are typically  implemented as distributed software-based systems that control embedded devices and field sensors. Yet to our knowledge, there has been no prior use of CE experiments on ICS. As such, relevant publications cover two distinct areas, namely \textit{(A)} vulnerability  and fault diagnosis in ICS, and \textit{(B)} current practices of CE in distributed systems. 

\subsection {Vulnerability Detection and Fault Testing in ICS}

There exist several techniques for vulnerability detection and failure testing in software-based distributed systems that share common characteristics with CE experiments. In software, existing approaches can be broadly classified into: 
\begin{enumerate}

\item Static and dynamic analysis methods including fuzzing techniques. These approaches identify issues inside the code of embedded components and industrial systems, either offline (static) or during run-time execution of software (dynamic).
\item Fault and attack injection on ICS components.  
\item Conditional testing approaches for the detection of system deviations, mostly implemented with invariant mining techniques.
\end{enumerate}

There is extensive research on static analysis of industrial software for fault detection. Recent approaches utilize various techniques, such as regression verification \cite{beckert2015regression}, model checking \cite{biallas2012arcade, canet2000towards}, or sequential function charts for the programming of the programmable logic controllers (PLCs). Similar approaches have extended these methods by including specifications of plant behavior converted to hybrid automata for verifying safety properties \cite{nellen2014cegar, nellen2016two}.

Concerning dynamic analysis approaches, there exists 
work that delves into fault testing of ICS components. 
Authors in \cite{guo2017symbolic} have presented how tools can generate test inputs through symbolic execution to detect program execution errors. VetPLC is another attempt that combines static and dynamic analysis presented in \cite{zhang2019towards}. Its goal is to verify real-world PLC code driven by events and detect safety violations in code. The work combines different types of analysis to understand causal relations among events in PLC code and quantitatively gauge temporal dependencies that are constrained by machine operations.

Published research has also used fuzz testing in ICS, with existing work targeting both industrial protocols and embedded systems. For example, authors in \cite{luo2019polar} were able to fuzz test function codes in ICS protocols and extract semantic information for vulnerability detection. Other approaches opted to fuzz input commands on firmware binaries extracted from ICS to evaluate the security of specific libraries inside ICS components \cite{muench2018you, tychalas2020}, or inject large number of attacks on network protocols and monitor target systems and their responses for unexpected behavior that might suggest the existence of vulnerabilities \cite{1633534, antunes2010vulnerability}.


Mining invariant values from ICS devices has also been in scope of recent research in fault detection of ICS software. Recent approaches aim to extract operational conditions from system logs via data mining. These conditional variables are then evaluated whether they hold true at some point in time during varying executions of ICS processes  \cite{beschastnikh2011leveraging}, \cite{ohmann2014behavioral}. Other approaches  mine such conditions from smart meters and medical devices, mostly for intrusion detection purposes \cite{aliabadi2017artinali}. Chen \textit{et al.}  extracted invariants from data traces of a water purification testbed to detect anomalies in system execution \cite{chen2018learning}. Other mixed approaches may also use invariants for vetting the source code of ICS components \cite{zhang2019towards}. 

CE experiments try to deal with the fact that functional charts and specifications cannot capture the scale and complexity of user behavior in modern distributed systems. As such, CE focuses on experimenting on the entire system under test as a single entity. Even though such approaches are not able to formally define function executions, they have shown great promise in uncovering unwanted states of execution that were previously unknown and not depicted in functional specifications \cite{basiri2016chaos, basiri2019automating}. CE experiments may also utilize fuzzing techniques for input, but they do not consider specific technical inputs or their fuzzed input does not target specific software or components. Instead, CE experiment view all ICS components and field devices as a single system and try to fuzz real-world, high-level input to observe what happens to the ICS boundary. 

Also, another difference between CE experiments and some of the aforementioned fault analysis and testing techniques lies on the fact that CE experiments do not follow a binary assertion logic (as is the case with invariants or static analysis). Most fault testing and assessment methods assign pre-considered restrictions and evaluate software-based systems against these precondition to assert them as true or false. Instead, CE experiments focus in creating new execution data and detect system states that were previously unknown to operators \cite{basiri2016chaos}.

\subsection{Existing CE Approaches}

Concerning existing CE experiments, the most famous case is Netflix, where engineers have taken CE as an approach to experimentally verify distributed systems' reliability. That is achieved either by changing the boundary state of components and analyzing system behavior using an internal service that was created specifically for his purpose (Chaos Monkey) \cite{basiri2016chaos}, or by introducing false injections at the boundary of Netflix micro-services \cite{basiri2019automating}. Chaos Monkey was also used in \cite{chang2015}, where researchers proposed a balanced use of the service to inject variable degrees of failure into the network without disconnecting it and assess it against network invariant metrics. 

Large-scale Java applications have also been known to utilize CE for testing complex distributed software \cite{zhang2019chaos}. Other implementations of the CE principles aimed to automatically inject system failures and errors into a containerized application to evaluate the overall system resilience under unknown states of operation \cite{simonsson2019}. Researchers in \cite{Przystalka2015} detected faults in industrial network data using recurrent neural networks that incorporate CE principles to improve the efficiency of the tuning procedure. Neural networks and CE was also proposed in \cite{aihara2002chaos} as a means for generating unwanted states in nonlinear electrical circuits. Authors aimed to discover vulnerabilities on large, complex systems, such as the U.S. power grid. 

Other research has focused on using CE for analyzing the execution states of Infrastructure-as-a-Service (IaaS) cloud platforms. For example, CloudStrike implemented the principles of CE by injecting faults to cloud resources to experiment on cybersecurity breaches caused by human errors and misconfigured resources  \cite{torkura2019security, torkura2020CloudStrike}. 

%% file: 3-cemodel_PJV.tex
\section{A Generic Chaos Engineering (CE) Model}\label{s:model}

The cyber-physical model of ICS can be seen as a model of a nonlinear system in the following state-space form: 
\begin{align}
\dot{x}(t) &=g(x(t), u(t), \delta(t)) \label{x} \\
\Delta y(t) &=h(x(t), u(t), \delta(t))
\label{y}
\end{align}
where the state $x(t) \in \mathbb{R}^{n}$ with $x(0)=x_{0}$, the measurement output deviations $\Delta y(t) \in \mathbb{R}^{m}$, the control input $u(t) \in \mathbb{R}^{m}$ (e.g., could represent set-points for states regulation), and $\mathcal{N}=$ $\{1, \cdots, n\}$ represents the index set of the ICS components.  $\delta(t) \in \mathbb{R}^{n_{\delta}}$ denotes the piecewise vector of constant 
signals including disturbances, measurement noise, and unknown control parameters. 
The model in Eqs. \eqref{x}, \eqref{y} may describe water treatment facility, a desalination plant, a microgrid or any other cyber-physical ICS, and might have been derived under suitable regularity constraints from a more general differential-algebraic model \cite{8025384}.  

\subsubsection{Steady state definition in ICS environments}

One major requirement of any CE experiment is to determine the system's steady state, i.e., that state in which the performance of the system is contributing towards maintaining a particular system property in a specific pattern or within a specific range. The objective in this part is the model development of the system in such a way to be able to describe the steady states according to the anticipated conditions of the system metrics. In the scenario of ICS, such metrics could be the level of sodium hydroxide in a water treatment plant \cite{oxfordus}.

\begin{definition}[Steady State Stability and Synchronization]\label{def:steadystatestability} There exist domains $\mathcal{V} \subseteq \mathbb{R}^{m} \times \mathbb{R}^{n_{\delta}}$ and $\mathcal{W} \subseteq \mathbb{R}^{n}$ for which:  

(1) $g$ and $h$ are Lipschitz continuous functions on $\mathcal{W} \times \mathcal{V}$,

(2) A Lipschitz continuous function $\phi_{w}$ on $\mathcal{W}$ is differentiable $\phi_{w}: \mathcal{V} \rightarrow \mathcal{W}$ and  $\forall (u, \delta) \in \mathcal{V}$ satisfies $0=g\left(\phi_{w}(u, \delta), u, \delta\right)$, i.e.,  each pair $(u, \delta) \in \mathcal{V}$ is unique on $\mathcal{W}$'s equilibrium state $\phi_{w}(u, \delta)$.

(3) There exist constants $c_{j}>0$, $j \in \mathbb{N}: j \in[1, 4]$, and a continuously differentiable function $f$ in $(x, u)$, $ f: \mathcal{W} \times \mathcal{V} \rightarrow \mathbb{R}_{\geq 0}, \quad(x,(u, \delta)) \mapsto f(x, u, \delta)$ for which $ \forall x \in \mathcal{W}$ and $ \forall (u, \delta) \in \mathcal{V}$: 
$$
\begin{array}{c}
c_{1}\left\|x-\phi_{w}(u, \delta)\right\|_{2}^{2} \leq f(x, u, w) \leq c_{2}\left\|x-\phi_{w}(u, \delta)\right\|_{2}^{2} \\
\nabla_{x} f(x, u, w)^{\top} f(x, u, \delta) \leq-c_{3}\left\|x-\phi_{w}(u, \delta)\right\|_{2}^{2} \\
\left\|\nabla_{u} f(x, u, \delta)\right\|_{2}<c_{4}\left\|x-\phi_{w}(u, \delta)\right\|_{2}
\end{array}
$$


\noindent (1)-(3) essentially present a singular perturbation problem in which relaxations and other variants exist, and (3) in particular, is a Lyapunov function determining exponential stability of  $\phi_{w}(u, \delta) \in \mathcal{W}$ \cite{phat2007exponential}.

(4) The input-to-output equilibrium mapping $\Delta \bar{y}(u, \delta)=h\left(\phi_{w}(u, \delta), u, w\right): \mathcal{V} \rightarrow \mathbb{R}^{m}$  has the form of:
\begin{align}
\begin{split}
 \Delta \bar{y}(u, \delta)&=h\left(\phi_{w}(u, \delta), u, w\right)\\
 =&({1}/{\gamma})* \mathbbm{1}_{m}*\left(\mathbbm{1}_{m}^{\top} \bar{u}-d\right)   
 \label{equlibrium}
\end{split}
\end{align}
where $\gamma>0$, $\mathbbm{1}_{m}$ representing the indicator function of size $m$, and $d_{cu} \in \mathbb{R}$ is an unmeasured constant disturbance.

\noindent  (4) shows that in steady state conditions, the synchronization of output measurements is possible with deviations being the same at all ICS components. The steady state of an output state can be seen as $\mathbbm{1}_{m}^{\top} u-d$, e.g., the difference of load demand and generation control inputs in a power grid model. The linear span of the indicator function is chosen due to the objective of having finite linear combinations as indicator step functions on arbitrary intervals. 
\end{definition}

In ICS, system operators or automated functionalities of the ICS allocate actions to the actuating elements of the systems in an effort to operate across nominal system setpoint metrics. The target setpoints can be determined in a minimization-type of problem:
\begin{align}
\begin{array}{cl}
\underset{{u}_{h} \in \mathbb{R}^{2}}{\operatorname{minimize}} & z({u}_{h}):=\sum_{j \in  \mathrm{V}} z_{j}\left({u}_{h}{_{j}}\right) \\
\text {s.t.} & 0=\mathbbm{1}_{m}^{\top} u-d_{cu}
\end{array}
\label{minimization}
\end{align}
where $j \in  \mathrm{V}$ and $|\mathrm{V}| = m$ is the set of the system nodes. Eq. \eqref{minimization} is assumed to be strictly feasible, i.e., the constraint is satisfied. $z_{j}: \mathcal{U}_{i} \rightarrow \mathbb{R}$ models the disutility of the $j_{\text{th}}$ ICS component of not satisfying system nominal demands, and encompasses a barrier function to enforce inequalities ${u}_{{h}_{j}} \in \mathcal{U}_{j}=\left({u}_{{l}_{j}}, {u}_{{h}_{j}}\right),$
where $-\infty \leq {u}_{{l}_{j}}<{u}_{{h}_{j}} \leq+\infty$.   The constraint of Eq. \eqref{minimization} ensures balance of system control inputs $\mathbbm{1}_{m}^{\mathrm{T}} u$ and unmeasured distrubances $d_{cu}$. In addition, by Eq.  \eqref{equlibrium}, Eq. \eqref{minimization} ensures that the steady state system deviations should lead to zero. An example metric following the last minimization problem could be defined as the frequency regulation setpoint of secondary frequency control of ICS in power grids. 

\subsubsection{Formulating hypotheses}

Once the metrics are determined and the steady state behavior of the ICS is understood, the experimental hypotheses can be defined. The objective is to determine what is expected as the result of an experiment, in terms of the system's steady state conditions, if we apply a set of diverse events into the ICS. Lemma \ref{distributedoptimalityconditions} relies on the minimization problem of Eq. \eqref{minimization} to determine under some assumptions the outcome of the experiment in terms of the target setpoints.

\begin{definition}[Direct Graph and Connectivity]\label{def:directgraph}
A sensor network of an ICS can be seen as a topological map represented by a directed graph $\mathcal{G}=(\mathrm{V}, \mathrm{E}, \mathrm{A})$ with nodes $\mathrm{V}=\{1, \ldots, m\}$, edges $\mathrm{E} \subseteq \mathrm{V} \times \mathrm{V}$, and $\mathrm{A}=\left[a_{i j}\right] \in \mathbb{R}^{m \times m} $ is a weighted
adjacency matrix with non-negative adjacency elements, i.e., $a_{i j} \geq 0$, and $a_{i j}>0~\text{if and only if the edge}~(i, j) \in \mathrm{E}$. The neighbors of node $i$ are defined as $\mathcal{N}_{i} \triangleq\{j \in \mathrm{V}:(i, j) \in \mathrm{E}\}$.
\end{definition}

\begin{definition}[Laplacian Matrix]\label{def:laplacian}
The Laplacian matrix $\mathrm{L} \in \mathbb{R}^{m \times m}$ of the graph $\mathcal{G}$ of Definition \ref{def:directgraph} is defined by its elements given by:
$$
L_{i j}=\left\{\begin{array}{ll}
-a_{i j} & \text { if } i \neq j \\
\sum_{k \neq i} a_{i k} & \text { if } i=j
\end{array}\right.
$$
Every row sum of $\mathrm{L}$ is zero. As for its eigenvalues $\lambda$, by definition there exist $\lambda_j = 0$ with right-eigenvector $\mathrm{X_{R_j}} = \mathbbm{1}_{m}$ and the rest $\lambda_i$, $i \neq j$, have positive real part \cite{7810500}. 
\end{definition}

\begin{definition}[Globally Reachable Node]\label{def:reachablenode}
A node $v_{i} \in \mathrm{V}$ is a globally reachable node, if there exists a path (i.e., a combination of edges in $\mathrm{E}$) from it to every other nodes in $\mathcal{G}$.  
\end{definition}

The next lemma shows an important property of Laplacian $\mathrm{L}$. 
\begin{lemma}\label{laplacianeigen}
The graph $\mathcal{G}$ has a globally reachable node if and only if the Laplacian $\mathrm{L}$ of $\mathcal{G}$
has a simple zero eigenvalue. In this scenario, the left-eigenvector $\mathrm{X_{L_j}} \in \mathbb{R}^{m}$ of $\mathrm{L}$ associated with the simple $\lambda_j = 0$ is $\mathrm{X_{L_j}} \geq 0$, and $\mathrm{X_{L_j}} >0$ if and only if node $v_{j} \in \mathrm{V}$ is globally reachable.
\end{lemma}

\begin{lemma}\label{distributedoptimalityconditions}
Consider Definition \ref{def:directgraph} of a sensor network of an ICS represented by a weighted directed graph $\mathcal{G}=(\mathrm{V}, \mathrm{E}, \mathrm{A})$ with Laplacian matrix $\mathrm{L}$, and let such graph to follow also Definition \ref{def:reachablenode}, i.e., include a globally reachable node and the left-eigenvector $\mathrm{X_{L_j}} \in \mathbb{R}_{\geq 0}^{m}$ of $\mathrm{L}$ to correspond to its simple eigenvalue $\lambda_j = 0$. Let $\Delta \bar{y}$ follows the form of Eq. \eqref{equlibrium} and ${u}_{h} \in \mathbb{R}^{m}$, then for the case of a diagonal matrix $D \succeq 0$ for which $\mathrm{X_{L_j}}^{\top} D \mathbbm{1}_{m}>0$, there exist the following equivalent statements: 
(1) ${u}_{h}$ is the unique optimal solution of Eq. \eqref{minimization}; and 
(2) the vector $q_h \in \operatorname{span}\left(\mathbbm{1}_{m}\right)$ is unique and satisfies:
\begin{subequations}
\begin{align}
0=D \Delta \bar{y}(u_h, \mathrm{X_{L_j}})+\mathrm{L} q_h \label{5a}\\
{u}_h=\nabla \cdot z^{*}(q_h)\label{5b}
\end{align}
\end{subequations}
where $z^{*}(q)=\sum_{j \in \mathrm{V}} z_{j}^{*}\left(q_{j}\right)$ is the conjugate of $z.$

\end{lemma}

The proofs of Lemma \ref{laplacianeigen} and Lemma \ref{distributedoptimalityconditions} are omitted due to space requirements and can be found in literature
\cite{bullo2009distributed, 7810500}.

The hypotheses in CE experiments are typically in the form of ``the events we are injecting into the system will not cause the system’s behavior to change from steady state'' \cite{rosenthal2017chaos}. Since the target metric setpoints can be determined via the optimization of Eq. \eqref{minimization}, and since according to Lemma \ref{distributedoptimalityconditions}, ${u}_h$ is the unique primal optimizer of Eq. \eqref{minimization}, then we can conclude that such formulation leads to a close-form solution not causing the system behavior to deviate from the steady state conditions. The close-form solution enables the analytical expression with a number of known functions in order to follow the concept of a well-defined steady behavior and hypotheses, in comparison with a problem solved with  a numerical solution which would be only an approximate.


\subsubsection{Varying ICS process events}

In any system, and therefore in any ICS, unpredictable events attributed to hardware or software faults/malfunctions, adverse malicious data, extreme operating conditions (e.g., high temperatures), etc. can lead to certain system conditions in which can further lead to system or sub-systems failures. In CE, the goal is to induce real-world events into the ICS which can be controlled. Following the formulation of the previous CE model, we follow the paradigm of linear distributed control or typically termed distributed averaging proportional integral (DAPI) control, often deployed in multi-agent networked systems in order to control system elements towards a consensus objective \cite{7990560, ge2021resilience}. Under this concept, the system components/agents in the distributed ICS (e.g., PLC devices) exchange information with other agents in order to take the proper control decisions (locally), often using linear PI (proportional integral) controls in an effort to minimize the agents' differences. 
If we consider the $\mathcal{N}$ agents of the ICS  to follow DAPI control, then the control dynamics of $u_{j}(t)$, $j \in \mathcal{N}$, can be represented as follows: 
\begin{equation}\label{dapi}
\dot{u}_{j}(t) =\sum_{\ell \in \mathcal{N}, \ell \neq j} a_{j \ell}\left(x_{\ell}(t)-x_{j}(t)\right)+\beta_{j}(t) 
\end{equation}
\noindent  $a_{i k}$ follows Definition \ref{def:directgraph} and it is a binary index of the unidirectional connectivity: if $a_{i k} = 1$, node $i$ can receive data from $k$, and  if $a_{i k} = 0$ there is no unidirectional communication from $k$ to $i$. $\beta_{j}(t)$ can be seen as the bias term to balance ICS operating set-points.

Let $C$ be a diagonal matrix to include the coefficients of the local measurement $C_{j}$ with $j=\{1, \ldots, n\}$. The differentiation of the local measurement data from other data sent from other system agents (remote) can lead to re-writing Eq. \eqref{dapi} as follows: 
\begin{equation}\label{dapiorganized}
\dot{u}_{j}(t) ={-C_{j} x_{j}(t)}+{\sum_{\ell \in \mathcal{N}, \ell \neq j} a_{j \ell} x_{\ell}(t)}+{\beta_{j}(t)}
\end{equation}
\noindent in which the terms represent the local information, the remote information, and the system bias, respectively. Following Definition \ref{def:directgraph} of the adjacency matrix representing the remote connectivity index of the ICS network, we can organize Eq. \eqref{dapiorganized} as: 
\begin{equation}\label{dapivector}
  \dot{u}(t)=(-C+A) x(t)+\beta(t)=-Q x(t)+\beta(t),  
\end{equation}
\noindent where $Q=C-A$. According to \cite{1333204}, and because  $Q$ is essentially a positive semi-definite matrix with every row sum being zero, the system has a steady-state equilibrium in consensus based on the control updates of Eq. \eqref{dapiorganized} if the ICS network agents form an undirected and connected graph. When $\beta=0$, any system equilibrium is in a consensus: by $x_{j}=x_{i}, j, i \in \mathcal{N}$.

\subsubsection{Run experiments in ICS production \& experiment automation}

A major difference of CE compared to accepted forms of security testing is the automation  within the system's production environment as well as the focus on the overall system behavior. Security automation drives continual supervision of the system, which especially in ICS is required due to continuous system changes, e.g., valves pressure, poisoning data, etc. We cannot be aware \textit{a priori} which conditions to the production environment will change the results of a CE test. In this experimentation stage, ideally, the tests need to be tested on the production environment. However, if that is not possible, experiments must be examined in testing conditions as close as possible to the production environment in order to reduce risks in terms of experiments validity  increased reliability, and provide enhanced insight into the performance of the experiments in terms of results confidence.

\subsubsection{Minimize the Blast Radius}

CE experiments should contribute in minimizing the blast radius of adverse events to the overall system operation, i.e., limit the range and severity of injected events to the system. A resilient ICS resulted from the CE experiments should reduce to the maximum the effects of adverse events. Such events, via the continuous CE experimentation, could be fully characterized providing full understanding of the initial conditions that drove them. Thus, it is of paramount importance to develop probabilistic models for asset availability, accessibility, and usability during and succeeding a major event. The task is complex and is dependent on multiple, interacting and often counterintuitive correlations between variables. 

%% file: 4-usecase.tex
\section{Use-Case Scenario}\label{s:use-case}

In this section, we introduce a use-case for CE on ICS that realizes the generic CE model presented in Section \ref{s:model} on a real-world case study. The work is based on the industrial process control model of the Tennessee Eastman Challenge (TEC) \cite{downs1993plant}. TEC defines a real-world industrial process control system (IPCS) for continuous chemical manufacturing plants that consists of five units: a reactor, a product condenser, a vapor-liquid separator, a recycle compressor, and a product stripper for producing the desired end products. The model has 12 actuators (manipulated variables) for control and 41 sensors (measured inputs) for monitoring. The model's ICS process produces two products from four input reactants. 

\begin{figure}[t]
    \centering
    \includegraphics[width=0.7\columnwidth]{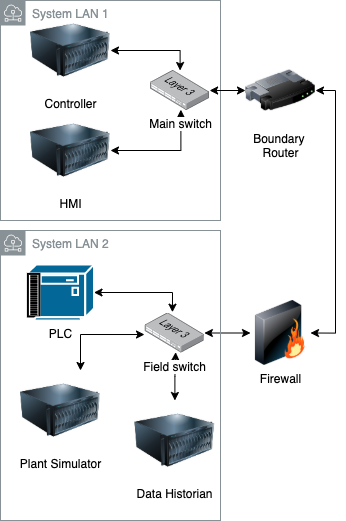}
    \caption{Network diagram of proposed testbed.}
    \label{fig:netdiagram}
\end{figure}

The presented use-case utilizes a variation of the IPCS of this testbed as set up by National Institute of Standards and Technology (NIST) to define an example where an adverse event can be detected through CE \cite{tang2017key}. This version of the model assumes a decentralized process controller for control loop continuous operation. It includes a PLC with industrial network protocol capability to provide communication between the field equipment and a PLC over a typical protocol (e.g., MODBUS). Field sensors send measured variable information to the PLC, and the PLC uses sensor inputs to compute desired values of the actuators. The IPCS network is segmented from the main network by a boundary router. Routing uses dynamic routing protocols to communicate with the main switch. The testbed utilizes two virtual networks, a main control room network (LAN 1) and the process operation environment (LAN 2). Network traffic between the two subnets is transferred through the boundary router. The operation environment consists of the plant's field sensor simulators, a PLC, and the data historian, while the control room consists of a human–machine interface
(HMI) and the controllers. The overall network architecture is depicted in Fig.  \ref{fig:netdiagram}.

\subsection{Step-by-step Description on a Real-world Environment}

For the purposes of introducing CE experiments in ICS, we could simplify the TEC testbed to only include one chemical reaction process that outputs \emph{product} $G$ using three input components. We could also restrict potential boundary conditions to the \emph{reactor temperature}, \emph{output yield of product} $G$, and \emph{input feed rate} of input materials such that the reaction rates of the reactants are a function of the reactor temperature. Product $G$ can be seen as the output of the chemical reaction of three input components: $u_{k}>0$, $k \in \mathbb{N}: k \in[1, 3]$ by sending vector signals $u = [u_1 … u_3]^T$ to the chemical process and receiving back sensor measurements $y = [y_1 … y_3]^T$. Sample nominal input values for the chemical process are described in Table \ref{tab:inputs}.


\begin{table}[t]
\setlength{\tabcolsep}{7pt}
\centering
\begin{tabular}{||c|l|c||}
\hline \hline
\textbf{Nominal Input} & \multicolumn{1}{c|}{\textbf{Description}} & \textbf{Value} \\ \hline \hline
$u_1$ & \begin{tabular}[c]{@{}l@{}}Chemical component A\\ feed setpoint to controller\end{tabular} & $0.25$ $kscmh$  \\ \hline
$u_2$ & \begin{tabular}[c]{@{}l@{}}Chemical component B\\ feed setpoint to controller\end{tabular} & $3686$ $kg/h$ \\ \hline
$u_3$ & \begin{tabular}[c]{@{}l@{}}Chemical component C\\ feed setpoint to controller\end{tabular} & $9.35$ $kscmh$ \\ \hline \hline
\end{tabular}
\caption{Sample nominal input values for chemical process production \cite{juricek2001identification}.}
\label{tab:inputs} 
\end{table}

\subsubsection{Steady state definition}

In order to setup a CE experiment on the aforementioned use-case, we first need to define the steady state of the system for that chemical process. We can describe this example as a stochastic state space model with additive noise such that it captures the evolution of the chemical process in the plant by describing wide ranges of time constants \cite{krishnamurthy2016partially}. Output deviations $\Delta y(t)$ on the production of $G$ and input ranges $u_i(t)$ for states regulation are defined as members of a same set $\mathbb{R}^{m}$. $\mathcal{V} \subseteq \mathbb{R}^{m} \times \mathbb{R}^{n_{\delta}}$ includes nominal input ranges for normal operating limits that induce acceptable states/members of set $\mathcal{W} \subseteq \mathbb{R}^{n}$. Composition measurements can be omitted to avoid a multi-rate sampling problem \cite{juricek2001identification}. 
The original publications that introduced the TEC testbed list the plant’s process nominal input demands as ranges of input values for equipment protection \cite{downs1993plant, tang2017key}. Variable value ranges are restricted by existing operating constrains of the TEC's chemical process, within finite domains of operation. These ranges are presented in Table \ref{steadyMetrics}. Ranges define the normal operating limits as well as the process shutdown limits for all chemical processes. 
At each time $t$, the state $x_{j}(t)$ of the $j_{th}$ ICS component that participates in the chemical production process lies within the state space $\mathbb{R}^{n}$, which is represented by the normal and shutdown values of five process variables (e.g., reactor temperature ranges). 

\begin{table}[t]
\setlength{\tabcolsep}{5pt}
\centering
\begin{tabular}{||l|l|l|l|l||}
\hline \hline
 & \multicolumn{2}{l|}{\textbf{\begin{tabular}[c]{@{}l@{}}Normal\\ operating limits\end{tabular}}} & \multicolumn{2}{l||}{\textbf{\begin{tabular}[c]{@{}l@{}}Shutdown\\ operating limits\end{tabular}}} \\ \hline \hline
\textbf{Process variable} & \textbf{Low limit} & \textbf{High limit} & \textbf{Low limit} & \textbf{High limit} \\ \hline \hline
Reactor pressure & None & $2895$ $kPa$ & None & $3000kPa$ \\ \hline
Reactor level & \begin{tabular}[c]{@{}l@{}}$50$\%\\ ($11.8$ $m^3$)\end{tabular} & \begin{tabular}[c]{@{}l@{}}$100$\%\\ ($21.3$ $m^3$)\end{tabular} & $2.0$ $m^3$ & $24.0$ $m^3$ \\ \hline
\begin{tabular}[c]{@{}l@{}}Reactor \\ temperature\end{tabular} & None & $150^\circ C$ & None & $175^\circ C$ \\ \hline
\begin{tabular}[c]{@{}l@{}}Product\\ separator level\end{tabular} & \begin{tabular}[c]{@{}l@{}}$30$\%\\ ($3.3$ $m^3$)\end{tabular} & \begin{tabular}[c]{@{}l@{}}100\%\\ ($9.0$ $m^3$)\end{tabular} & $1.0$ $m^3$ & $12.0$ $m^3$ \\ \hline
\begin{tabular}[c]{@{}l@{}}Stripper\\ base level\end{tabular} & \begin{tabular}[c]{@{}l@{}}$30$\%\\ ($3.5$ $m^3$)\end{tabular} & \begin{tabular}[c]{@{}l@{}}$100$\%\\ ($6.6$ $m^3$)\end{tabular} & $1.0$ $m^3$ & $8.0$ $m^3$ \\ \hline \hline
\end{tabular}
\caption{Process operating constraints \cite{downs1993plant}.}
\label{steadyMetrics}
\end{table}

A CE experiment on the TEC testbed can utilize operational metrics such as the \emph{throughput rate}, \emph{input feed rate}, or \emph{output yield} of a checmical process to monitor for potential deviations $y_{t}$ that capture the execution of the chemical process at the system boundary. These metrics, depicted in Table \ref{tMetrics}, can act as primary indicators of the reliability of the industrial process and characterize the hypothesis of the steady state behavior. 
The operational metrics capture the ICS chemical process's availability at the system boundary, and not technical metrics such as reactor temperature or pressure levels since this is not in line with CE experimentation.

\subsubsection{Hypothesis formulation}

Steady state conditions allow for deviations within the set of normal operating states as depicted in Table \ref{steadyMetrics} for all process variables of the ICS components. For example, for the aforementioned chemical process of product $G$, the steady state can be seen as the difference of the dependent process variable values (e.g., the reactor temperature or any operational metric) and the chemical process's control inputs for the production of $G$.

Following Definition \ref{def:directgraph}, all networked systems along with the plant simulator that participate in this chemical process can be represented in a weighted, directed graph that has at least one globally reachable node (i.e., its left-eigenvector $\mathrm{X_{L_j}} \geq 0$, and $\mathrm{X_{L_j}} >0$). We assume agents of the TEC architecture to follow DAPI control with state vectors $x_{j}(t)$ seen as a linear combination of control inputs $u_{j}(t)$, $j \in \mathcal{N}$ and outputs $\Delta y(t)$ over time ${t}$ with a sampling data period of $\Delta t = 1$ min, computed from data since the initial startup of the plant’s process. 

For the purposes of the CE experiment, we form the hypothesis that the output of the ``\emph{output yield}'' metric will remain within acceptable bounds, given any set of linear combinations of control inputs \cite{rosenthal2017chaos}. Output can be mapped as a single output $y_{t}$ vector in $\mathbb{R}^{m}$. The ``\emph{output yield}'' metric is chosen as an example. Different types of key performance indicators (KPIs) can be chosen, based on operator knowledge and on the characteristics of the system under test, e.g., manufacturing process performance KPIs, computing resources performance KPIs, OPC data exchange performance KPIs, etc.

\subsubsection{Varying ICS process events}

\begin{table}[t]
\setlength{\tabcolsep}{6pt}
\centering
\begin{tabular}{||l|l||}
\hline \hline
\textbf{System Boundary metric} & \textbf{{ Description}} \\ \hline \hline 
\textit{Throughput Rate} & \begin{tabular}[c]{@{}l@{}}Measures amount of end products\\ produced over a specific amount of time.\end{tabular} \\ \hline
\textit{Input Feed Rate} & \begin{tabular}[c]{@{}l@{}}Measures amount of input materials\\ consumed over a specific amount of time.\end{tabular} \\ \hline
\textit{Output Yield} & \begin{tabular}[c]{@{}l@{}}Measures percentage of the end product \\ produced at the output of the process.\end{tabular} \\ \hline \hline
\end{tabular}
\caption{Manufacturing process key performance indicators (KPIs) as boundary metrics \cite{tang2017key}.}
\label{tMetrics} 
\end{table}

The CE experiment will vary potential real-world events to affect the modeled steady state and then analyze its impact on the boundary metric ``\emph{output yield}'' as described in Table \ref{tMetrics}. These variations will mostly take the form of non-critical events that can still affect the overall ICS functionality and disrupt the ICS steady state. 

Agents in the used TEC closed-loop system theoretically ensure the reliability of the chemical process during the occurrence of all potential events. Still, when certain limits are crossed in critical and non-critical systems, ICS production and state of operation may not satisfy system nominal demands through the disutility $z({u}_{h})$ of an ICS component that participates in the chemical process to create product $G$, as defined in Eq. \ref{minimization}. Furthermore, close-loop processes may often be disrupted and fall back to alternate operations. In such events, ICS often utilize backup analog mechanisms (e.g., meters, alarms, etc.) to continue observing system states. Still, this behavior may not be desirable in the long-term since it may require additional time and effort to perform necessary monitoring or control functions, which may present risks due to reduced quality, safety, or efficiency of the system \cite{stouffer2011guide}. 

The proposed CE experiment focuses only on the ICS process for the production of end product $G$ as captured by the output yield metric over time periods ${t}$. The process evolves according to the state vector $x_{t}$ and inputs $u_{t}$. A list of such variations for the presented testbed can be quite diverse, including (but not restricted to) the following:

\begin{itemize}
\item Terminate/overload server instances (e.g., historian, HMI).
\item Inject latency into requests between the controller and the sensors.
\item Inject network errors between the controller and the actuator. 
\item Make a subnet unavailable (router failure).
\item Restart the PLC at a given point in the industrial process.
\item Fail a sensor.
\end{itemize}
This list includes sample input events chosen due to their ability to multiply the effect of the CE experiment and introduce fluctuations at a broad system level (in our case, as a significant deviation on the \textit{output yield rate} for the chemical process of product $G$) instead of granular, specific technical inconsistencies. We can effectively model these fluctuations as step changes in the chemical process using a simple linear model \cite{juricek2001identification}. Typical ICS experiments can last from 24 to 48 hours and capture information to calculate relevant boundary metrics over the IPCS potential nominal input values \cite{tang2017key}. 
As depicted in Eq. \eqref{eq:u0}, inputs can be seen as a sequence of signals, where $u^{0}_{i}(t)$ is the nominal input for the base operating mode and reference inputs given in Table \ref{tab:inputs}. Following a similar case study \cite{juricek2001identification}, we can perturb manipulated input variables as three-level sequences (e.g.,  pseudo random binary sequences (PRBS)) to generate deterministic input signals $u_{j}({t})$ and reduce the effect of non-linearities on the resulting linear model \cite{godfrey1993perturbation}. 
\begin{align}
    u^{0}_{i}(t) = \left\{\begin{array}{rl}
    \begin{split}
        u_{i}^{0}(t), \quad \quad \quad \quad \quad  t < t_{0}
    \end{split}\\
    \begin{split}
        u_{i}^{0}(t) + \Delta u_{i}(t), \quad t_{0} \leq t < (t_{0} + 24h)
    \end{split}\\
    \begin{split}
        u_{i}^{0}(t) - \Delta u_{i}(t), \quad t_{0} +24h \leq t < (t_{0} + 48h)
    \end{split}\\
    \begin{split}
        u_{i}^{0}(t), \quad \quad \quad \quad \quad  t \geq (t_{0} + 48h)
    \end{split}
    \end{array}\right.
    \label{eq:u0}
\end{align}

\underline{\textit{Example use-case}}: The PLC scans for signals gathered from the sensors monitoring the reactor. If there is an increase in reactor temperature, then the PLC will communicate with the actuator to initiate commands for decreasing the temperature at the reactor.  An adverse event introduces lots of traffic at the protocol level and causes a congestion failure at the boundary router. Field metrics and scans do not reach the control room in timely order, although the system does not yet detect any disconnection to the field. As such, the reactor temperature rises and the output yield of product $G$ decreases to unacceptable levels. From a business process perspective, this event sums up to whether the generator is able to reliably operate on any acceptable steady state and deliver its services in any case scenario. To analyze such boundary system states from a functional perspective, we map the system's operational ``output yield'' output for product $G$ that reliably sets a vector of measured outputs to monitor for states that do not satisfy system nominal demands.

\subsubsection{Run experiments in ICS production \&  analyze  results}

During CE experimentation, we vary control inputs in an non-binary assertion logic. By following Eq. \eqref{5b}, the CE experiment aims to measure the divergence (whether positive or negative) from any steady state operation of the system towards a state that does not satisfy system nominal demands.

In the model test-case, unpredictable events on networked components should be used as input to the CE experiment. These events must ideally follow a non-deterministic approach that still conforms to real-world execution conditions of the chemical process for producing product $G$. For example, an experiment can utilize software or hardware malfunctions, or extreme input to controllers but obviously cannot generate events that can never happen on the real-world (e.g., assume different ICS components or changes in the system's structure). 
If the CE experiment is run on a real production environment, experimentation must implement safety measures that will enable operators to recover individual ICS components to previously defined steady states \cite{torkura2019security}. This safety practice in CE experiments is in line with best practices by NIST, which specifically state that exhaustive testing of systems is essential to support availability of services \cite{jillepalli2017security}. This may include saving and restoring configuration data as a form of ICS component state, or even utilizing in-memory data to mitigate effects in erroneous system states. 
For all intends and purposes, if semi-simulated environments are used for running CE experiments, then each recorded adverse event should be externally validated to assure operators that all unwanted states detected will generalize to the real system, or is simply a product of experimentation on the testbed that will not reflect to real-world systems \cite{rosenthal2017chaos}.

\subsection{Mitigating Operational Risks}

\begin{figure}[t]
    \centering
    \includegraphics[width=\columnwidth]{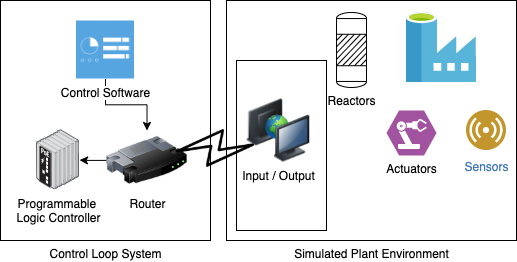}
    \caption{A hardware-in-the-loop (HIL)-based ICS testbed.}
    \label{fig:hil}
\end{figure}

Traditionally, CE experiments are purposely run in production to analyze the behavior of the entire real-world system and capture potential third-party influence issues (e.g., inputs, other systems, library dependencies, etc.). Still, the ICS environment is heavily dependent on constant availability, which renders such approaches too risky. Instead, CE events can be simulated using a HIL implementation of the aforementioned model. Such testbeds are efficient when challenging the resiliency of ICS against highly complex scenarios \cite{grega1999hardware}, since they integrate the complexity of the real-world system by adding the actual control system and simulating the plant equipment. This way, they mimic real-world systems without field devices and machinery, while utilizing real hardware for the control loop. In HIL testbeds, the software layer must reflect how each added component to the ICS increases the attack surface \cite{mclaughlin2016cybersecurity, 9351954}. An example layout of HIL testbeds in shown in Fig. \ref{fig:hil}.

CE experiments conducted on HIL testbeds can test actual control systems without having to experiment on real-world field devices and eliminating test risk. We can simulate failure events by selectively failing the aforementioned systems on the HIL testbed and analyze the resulting execution states of engaged hardware devices (e.g., sudden termination of the historian, or take a router offline). Since synthetic models cannot possibly capture the real-world complexity or randomness of events, it is important to utilize as many devices as possible on the HIL testbed. Ideally, only field machinery, actuators and sensors should be simulated, to avoid real-world impact.

%% file: 5-conclusions_PJV.tex
\section{Conclusions and Future Work}\label{s:conclusions}


Research has intensively explored systems and software testing, using various approaches such as execution path analysis, injection, fuzzing and conditional detection methods. Still, current work seems inadequate to capture the complexity of functionality, specifications and user behavior in modern distributed and industrial systems. In this paper, we adopt the concept of CE and present how it can be utilized for complex CPS, and specifically, ICS environments in which adverse events can affect the operating system states. We have examined the related work in the area and set up a formal CE model of nonlinear ICS that generalizes over the CE experimentation steps. We have also described a use-case scenario based on the TEC chemical process. In our future work, we aim to develop a resilience assessment framework in a methodological approach that will utilize the concept of CE to quantify the resilience of industrial CPS. The goal is not only to expand the theoretical and experimentation contribution of this work, but also develop automated restoration policies against classes of adverse events which can be utilized in practical systems. Furthermore, we plan to identify necessary and sufficient conditions which can ensure steady state system operation under stealthy cyberattacks and evaluate them in a realistic HIL testbed.

%% file: main.bbl
\begin{thebibliography}{10}
\providecommand{\url}[1]{#1}
\csname url@samestyle\endcsname
\providecommand{\newblock}{\relax}
\providecommand{\bibinfo}[2]{#2}
\providecommand{\BIBentrySTDinterwordspacing}{\spaceskip=0pt\relax}
\providecommand{\BIBentryALTinterwordstretchfactor}{4}
\providecommand{\BIBentryALTinterwordspacing}{\spaceskip=\fontdimen2\font plus
\BIBentryALTinterwordstretchfactor\fontdimen3\font minus
  \fontdimen4\font\relax}
\providecommand{\BIBforeignlanguage}[2]{{%
\expandafter\ifx\csname l@#1\endcsname\relax
\typeout{** WARNING: IEEEtran.bst: No hyphenation pattern has been}%
\typeout{** loaded for the language `#1'. Using the pattern for}%
\typeout{** the default language instead.}%
\else
\language=\csname l@#1\endcsname
\fi
#2}}
\providecommand{\BIBdecl}{\relax}
\BIBdecl

\bibitem{paulo1}
P.~Verissimo, M.~Correia, N.~F. Neves, and P.~Sousa, ``Intrusion-resilient
  middleware design and validation,'' \emph{Information Assurance, Security and
  Privacy Services}, vol.~4, pp. 615--678, 2009.

\bibitem{8088656}
T.~R. {McJunkin} and C.~G. {Rieger}, ``Electricity distribution system
  resilient control system metrics,'' in \emph{2017 Resilience Week (RWS)},
  2017, pp. 103--112.

\bibitem{9351954}
I.~{Zografopoulos}, J.~{Ospina}, X.~{Liu}, and C.~{Konstantinou},
  ``Cyber-physical energy systems security: Threat modeling, risk assessment,
  resources, metrics, and case studies,'' \emph{IEEE Access}, vol.~9, pp.
  29\,775--29\,818, 2021.

\bibitem{koulamas2018cyber}
C.~Koulamas and A.~Kalogeras, ``Cyber-physical systems and digital twins in the
  industrial internet of things [cyber-physical systems],'' \emph{Computer},
  vol.~51, no.~11, pp. 95--98, 2018.

\bibitem{yu2020resilient}
F.~Yu, Y.~Hu, T.~Zhang, and Y.~Jin, ``Resilient distributed estimator with
  information consensus for cps security,'' in \emph{2020 IEEE 38th
  International Conference on Computer Design (ICCD)}.\hskip 1em plus 0.5em
  minus 0.4em\relax IEEE, 2020, pp. 41--44.

\bibitem{9496085}
C.~Konstantinou and O.~M. Anubi, ``Resilient cyber-physical energy systems
  using prior information based on gaussian process,'' \emph{IEEE Transactions
  on Industrial Informatics}, pp. 1--1, 2021.

\bibitem{8255544}
R.~{Nateghi}, ``Multi-dimensional infrastructure resilience modeling: An
  application to hurricane-prone electric power distribution systems,''
  \emph{IEEE Access}, vol.~6, pp. 13\,478--13\,489, 2018.

\bibitem{8221708}
C.~{Konstantinou}, M.~{Sazos}, A.~S. {Musleh}, A.~{Keliris}, A.~{Al-Durra}, and
  M.~{Maniatakos}, ``Gps spoofing effect on phase angle monitoring and control
  in a real-time digital simulator-based hardware-in-the-loop environment,''
  \emph{IET Cyber-Physical Systems: Theory Applications}, vol.~2, no.~4, pp.
  180--187, 2017.

\bibitem{miles2019learning}
R.~Miles, \emph{Learning Chaos engineering: discovering and overcoming system
  weaknesses through experimentation}.\hskip 1em plus 0.5em minus 0.4em\relax
  O'Reilly Media, 2019.

\bibitem{basiri2016chaos}
A.~{Basiri}, N.~{Behnam}, R.~{de Rooij}, L.~{Hochstein}, L.~{Kosewski},
  J.~{Reynolds}, and C.~{Rosenthal}, ``Chaos engineering,'' \emph{IEEE
  Software}, vol.~33, no.~3, pp. 35--41, 2016.

\bibitem{rosenthal2017chaos}
C.~Rosenthal, L.~Hochstein, A.~Blohowiak, N.~Jones, and A.~Basiri, \emph{Chaos
  Engineering: Building Confidence in System Behavior Through
  Experiments}.\hskip 1em plus 0.5em minus 0.4em\relax O'Reilly Media, 2017.

\bibitem{beckert2015regression}
B.~Beckert, M.~Ulbrich, B.~Vogel-Heuser, and A.~Weigl, ``Regression
  verification for programmable logic controller software,'' in
  \emph{International Conference on Formal Engineering Methods}.\hskip 1em plus
  0.5em minus 0.4em\relax Springer, 2015, pp. 234--251.

\bibitem{biallas2012arcade}
S.~Biallas, J.~Brauer, and S.~Kowalewski, ``Arcade. plc: A verification
  platform for programmable logic controllers,'' in \emph{2012 Proceedings of
  the 27th IEEE/ACM International Conference on Automated Software
  Engineering}.\hskip 1em plus 0.5em minus 0.4em\relax IEEE, 2012, pp.
  338--341.

\bibitem{canet2000towards}
G.~Canet, S.~Couffin, J.-J. Lesage, A.~Petit, and P.~Schnoebelen, ``Towards the
  automatic verification of plc programs written in instruction list,'' in
  \emph{Smc 2000 conference proceedings. 2000 ieee international conference on
  systems, man and cybernetics.'cybernetics evolving to systems, humans,
  organizations, and their complex interactions'(cat. no. 0}, vol.~4.\hskip 1em
  plus 0.5em minus 0.4em\relax IEEE, 2000, pp. 2449--2454.

\bibitem{nellen2014cegar}
J.~Nellen, E.~{\'A}brah{\'a}m, and B.~Wolters, ``A cegar tool for the
  reachability analysis of plc-controlled plants using hybrid automata,'' in
  \emph{Workshop on Formal Methods Integration}.\hskip 1em plus 0.5em minus
  0.4em\relax Springer, 2014, pp. 55--78.

\bibitem{nellen2016two}
J.~Nellen, K.~Driessen, M.~Neuh{\"a}u{\ss}er, E.~{\'A}brah{\'a}m, and
  B.~Wolters, ``Two cegar-based approaches for the safety verification of
  plc-controlled plants,'' \emph{Information Systems Frontiers}, vol.~18,
  no.~5, pp. 927--952, 2016.

\bibitem{guo2017symbolic}
S.~Guo, M.~Wu, and C.~Wang, ``Symbolic execution of programmable logic
  controller code,'' in \emph{Proceedings of the 2017 11th Joint Meeting on
  Foundations of Software Engineering}, 2017, pp. 326--336.

\bibitem{zhang2019towards}
M.~Zhang, C.-Y. Chen, B.-C. Kao, Y.~Qamsane, Y.~Shao, Y.~Lin, E.~Shi, S.~Mohan,
  K.~Barton, J.~Moyne \emph{et~al.}, ``Towards automated safety vetting of plc
  code in real-world plants,'' in \emph{2019 IEEE Symposium on Security and
  Privacy (SP)}.\hskip 1em plus 0.5em minus 0.4em\relax IEEE, 2019, pp.
  522--538.

\bibitem{luo2019polar}
Z.~Luo, F.~Zuo, Y.~Jiang, J.~Gao, X.~Jiao, and J.~Sun, ``Polar: Function code
  aware fuzz testing of ics protocol,'' \emph{ACM Transactions on Embedded
  Computing Systems (TECS)}, vol.~18, no.~5s, pp. 1--22, 2019.

\bibitem{muench2018you}
M.~Muench, J.~Stijohann, F.~Kargl, A.~Francillon, and D.~Balzarotti, ``What you
  corrupt is not what you crash: Challenges in fuzzing embedded devices.'' in
  \emph{NDSS}, 2018.

\bibitem{tychalas2020}
D.~{Tychalas} and M.~{Maniatakos}, ``Iffset: In-field fuzzing of industrial
  control systems using system emulation,'' in \emph{2020 Design, Automation
  Test in Europe Conference Exhibition (DATE)}, 2020, pp. 662--665.

\bibitem{1633534}
N.~Neves, J.~Antunes, M.~Correia, P.~Verissimo, and R.~Neves, ``Using attack
  injection to discover new vulnerabilities,'' in \emph{International
  Conference on Dependable Systems and Networks (DSN'06)}, 2006, pp. 457--466.

\bibitem{antunes2010vulnerability}
J.~Antunes, N.~Neves, M.~Correia, P.~Verissimo, and R.~Neves, ``Vulnerability
  discovery with attack injection,'' \emph{IEEE Transactions on Software
  Engineering}, vol.~36, no.~3, pp. 357--370, 2010.

\bibitem{beschastnikh2011leveraging}
I.~Beschastnikh, Y.~Brun, S.~Schneider, M.~Sloan, and M.~D. Ernst, ``Leveraging
  existing instrumentation to automatically infer invariant-constrained
  models,'' in \emph{Proceedings of the 19th ACM SIGSOFT symposium and the 13th
  European conference on Foundations of software engineering}, 2011, pp.
  267--277.

\bibitem{ohmann2014behavioral}
T.~Ohmann, M.~Herzberg, S.~Fiss, A.~Halbert, M.~Palyart, I.~Beschastnikh, and
  Y.~Brun, ``Behavioral resource-aware model inference,'' in \emph{Proceedings
  of the 29th ACM/IEEE international conference on Automated software
  engineering}, 2014, pp. 19--30.

\bibitem{aliabadi2017artinali}
M.~R. Aliabadi, A.~A. Kamath, J.~Gascon-Samson, and K.~Pattabiraman,
  ``Artinali: dynamic invariant detection for cyber-physical system security,''
  in \emph{Proceedings of the 2017 11th Joint Meeting on Foundations of
  Software Engineering}, 2017, pp. 349--361.

\bibitem{chen2018learning}
Y.~Chen, C.~M. Poskitt, and J.~Sun, ``Learning from mutants: Using code
  mutation to learn and monitor invariants of a cyber-physical system,'' in
  \emph{2018 IEEE Symposium on Security and Privacy (SP)}.\hskip 1em plus 0.5em
  minus 0.4em\relax IEEE, 2018, pp. 648--660.

\bibitem{basiri2019automating}
A.~{Basiri}, L.~{Hochstein}, N.~{Jones}, and H.~{Tucker}, ``Automating chaos
  experiments in production,'' in \emph{2019 IEEE/ACM 41st International
  Conference on Software Engineering: Software Engineering in Practice
  (ICSE-SEIP)}, 2019, pp. 31--40.

\bibitem{chang2015}
\BIBentryALTinterwordspacing
M.~A. Chang, B.~Tschaen, T.~Benson, and L.~Vanbever, ``Chaos monkey: Increasing
  sdn reliability through systematic network destruction,'' in
  \emph{Proceedings of the 2015 ACM Conference on Special Interest Group on
  Data Communication}, ser. SIGCOMM '15.\hskip 1em plus 0.5em minus 0.4em\relax
  New York, NY, USA: Association for Computing Machinery, 2015, p. 371–372.
  [Online]. Available: \url{https://doi.org/10.1145/2785956.2790038}
\BIBentrySTDinterwordspacing

\bibitem{zhang2019chaos}
L.~{Zhang}, B.~{Morin}, P.~{Haller}, B.~{Baudry}, and M.~{Monperrus}, ``A chaos
  engineering system for live analysis and falsification of exception-handling
  in the jvm,'' \emph{IEEE Transactions on Software Engineering}, pp. 1--1,
  2019.

\bibitem{simonsson2019}
J.~Simonsson, L.~Zhang, B.~Morin, B.~Baudry, and M.~Monperrus, ``Observability
  and chaos engineering on system calls for containerized applications in
  docker,'' 07 2019.

\bibitem{Przystalka2015}
P.~Przystałka and W.~Moczulski, ``Methodology of neural modelling in fault
  detection with the use of chaos engineering,'' \emph{Engineering Applications
  of Artificial Intelligence}, vol.~41, 05 2015.

\bibitem{aihara2002chaos}
K.~{Aihara}, ``Chaos engineering and its application to parallel distributed
  processing with chaotic neural networks,'' \emph{Proceedings of the IEEE},
  vol.~90, no.~5, pp. 919--930, 2002.

\bibitem{torkura2019security}
K.~A. {Torkura}, M.~I.~H. {Sukmana}, F.~{Cheng}, and C.~{Meinel}, ``Security
  chaos engineering for cloud services: Work in progress,'' in \emph{2019 IEEE
  18th International Symposium on Network Computing and Applications (NCA)},
  2019, pp. 1--3.

\bibitem{torkura2020CloudStrike}
------, ``Cloudstrike: Chaos engineering for security and resiliency in cloud
  infrastructure,'' \emph{IEEE Access}, vol.~8, pp. 123\,044--123\,060, 2020.

\bibitem{8025384}
K.~{Paridari}, N.~{O’Mahony}, A.~{El-Din Mady}, R.~{Chabukswar},
  M.~{Boubekeur}, and H.~{Sandberg}, ``A framework for attack-resilient
  industrial control systems: Attack detection and controller
  reconfiguration,'' \emph{Proceedings of the IEEE}, vol. 106, no.~1, pp.
  113--128, 2018.

\bibitem{oxfordus}
O.~Analytica, ``Us cyberattack underlines sub-national risks,'' \emph{Emerald
  Expert Briefings}, no. oxan-es.

\bibitem{phat2007exponential}
V.~N. Phat and P.~T. Nam, ``Exponential stability and stabilization of
  uncertain linear time-varying systems using parameter dependent lyapunov
  function,'' \emph{International Journal of Control}, vol.~80, no.~8, pp.
  1333--1341, 2007.

\bibitem{7810500}
J.~{Schiffer} and F.~{Dörfler}, ``On stability of a distributed averaging pi
  frequency and active power controlled differential-algebraic power system
  model,'' in \emph{2016 European Control Conference (ECC)}, 2016, pp.
  1487--1492.

\bibitem{bullo2009distributed}
F.~Bullo, J.~Cort{\'e}s, and S.~Martinez, \emph{Distributed control of robotic
  networks: a mathematical approach to motion coordination algorithms}.\hskip
  1em plus 0.5em minus 0.4em\relax Princeton University Press, 2009.

\bibitem{7990560}
D.~K. {Molzahn}, F.~{Dörfler}, H.~{Sandberg}, S.~H. {Low}, S.~{Chakrabarti},
  R.~{Baldick}, and J.~{Lavaei}, ``A survey of distributed optimization and
  control algorithms for electric power systems,'' \emph{IEEE Transactions on
  Smart Grid}, vol.~8, no.~6, pp. 2941--2962, 2017.

\bibitem{ge2021resilience}
P.~Ge, F.~Teng, C.~Konstantinou, and S.~Hu, ``A resilience-oriented
  centralised-to-decentralised framework for networked microgrids management,''
  \emph{arXiv preprint arXiv:2109.00245}, 2021.

\bibitem{1333204}
R.~{Olfati-Saber} and R.~M. {Murray}, ``Consensus problems in networks of
  agents with switching topology and time-delays,'' \emph{IEEE Transactions on
  Automatic Control}, vol.~49, no.~9, pp. 1520--1533, 2004.

\bibitem{downs1993plant}
J.~J. Downs and E.~F. Vogel, ``A plant-wide industrial process control
  problem,'' \emph{Computers \& chemical engineering}, vol.~17, no.~3, pp.
  245--255, 1993.

\bibitem{tang2017key}
C.~Tang and C.~Tang, \emph{Key performance indicators for process control
  system cybersecurity performance analysis}.\hskip 1em plus 0.5em minus
  0.4em\relax US Department of Commerce, National Institute of Standards and
  Technology, 2017.

\bibitem{juricek2001identification}
B.~C. Juricek, D.~E. Seborg, and W.~E. Larimore, ``Identification of the
  tennessee eastman challenge process with subspace methods,'' \emph{Control
  Engineering Practice}, vol.~9, no.~12, pp. 1337--1351, 2001.

\bibitem{krishnamurthy2016partially}
V.~Krishnamurthy, \emph{Partially observed Markov decision processes}.\hskip
  1em plus 0.5em minus 0.4em\relax Cambridge university press, 2016.

\bibitem{stouffer2011guide}
K.~Stouffer, J.~Falco, and K.~Scarfone, ``Guide to industrial control systems
  (ics) security,'' \emph{NIST special publication}, vol. 800, no.~82, pp.
  16--16, 2011.

\bibitem{godfrey1993perturbation}
K.~Godfrey, \emph{Perturbation signals for system identification}.\hskip 1em
  plus 0.5em minus 0.4em\relax Prentice Hall International (UK) Ltd., 1993.

\bibitem{jillepalli2017security}
A.~A. Jillepalli, F.~T. Sheldon, D.~C. de~Leon, M.~Haney, and R.~K.
  Abercrombie, ``Security management of cyber physical control systems using
  nist sp 800-82r2,'' in \emph{2017 13th International Wireless Communications
  and Mobile Computing Conference (IWCMC)}.\hskip 1em plus 0.5em minus
  0.4em\relax IEEE, 2017, pp. 1864--1870.

\bibitem{grega1999hardware}
W.~Grega, ``Hardware-in-the-loop simulation and its application in control
  education,'' in \emph{FIE'99 Frontiers in Education. 29th Annual Frontiers in
  Education Conference. Designing the Future of Science and Engineering
  Education. Conference Proceedings (IEEE Cat. No. 99CH37011}, vol.~2.\hskip
  1em plus 0.5em minus 0.4em\relax IEEE, 1999, pp. 12B6--7.

\bibitem{mclaughlin2016cybersecurity}
S.~McLaughlin, C.~Konstantinou, X.~Wang, L.~Davi, A.-R. Sadeghi, M.~Maniatakos,
  and R.~Karri, ``The cybersecurity landscape in industrial control systems,''
  \emph{Proceedings of the IEEE}, vol. 104, no.~5, pp. 1039--1057, 2016.

\end{thebibliography}
